# Understanding Emotional Hijacking in Metaverse

SYED ALI ASIF, PHILIP GABLE, and CHIEN-CHUNG SHEN, University of Delaware, USA
YAN-MING CHIOU, SRI, USA

## POSITION STATEMENT

Emotions are an integral part of being human, and experiencing a range of emotions is what makes life rich and vibrant. From basic emotions like anger, fear, happiness, and sadness to more complex ones like excitement and grief, emotions help us express ourselves and connect with the world around us. In recent years, researchers have begun adopting virtual reality (VR) technology to evoke emotions as realistically as possible and quantify the strength of emotions from the electroencephalogram (EEG) signals measured from the brain to understand human emotions in realistic situations better. This is achieved by creating a sense of presence in the virtual environment, the feeling that the user is there. For instance, [6] studied the excitement of a rollercoaster ride in VR, and [5] studied the fear of navigating in a VR cave.

While VR is being used as a tool to help study emotions, the advances of VR itself have facilitated the development of the Metaverse, a fully immersive virtual world where people, through their avatars, can interact with a virtual environment and with each other in real-time. Millions of people have already experienced some aspects of the Metaverse through popular online, multi-user VR games like Roblox and Fortnite. Though multi-user VR games are limited to their specific gaming environment and objectives, the Metaverse is designed to be open and can also include virtual marketplaces, social spaces, educational environments, and more. It is a digital universe where users can live, work, play, and socialize in a completely immersive way. Furthermore, while avatars in the existing Metaverse (and multi-user VR games) are cartoon-like figures, new technologies like Generative AI [9] and Neural Radiance Field (NeRF) [7] allow for the creation of realistic 3D avatars (from 2D photos). Also, full-body avatars, like Codec Avatar [4], represent a step towards making the Metaverse even more realistic. Moreover, the built-in eye and facial tracking capability in VR headsets lets the avatar express the genuine emotion of its user in real-time.

The Metaverse is envisioned as a 3D version of the (2D) Internet, where users can navigate through a variety of virtual spaces and engage in a range of activities, including socializing, gaming, shopping, and learning. Although versatile and exciting, the Metaverse may also present psychological challenges, such as the inability to distinguish between virtual and physical reality, leading to *emotional hijacking*. The recent Korean movie "Hee-Soo" [10] painted a vivid yet alarming picture of the potential negative impacts of the Metaverse, where a mother "re-united" with her deceased daughter in the Metaverse and gradually lost her own sense of reality.

Emotional hijacking is a term coined by Daniel Goleman [3] to describe a state where emotions overwhelm the brain's ability to function. Initially, Goleman described it as an "amygdala hijack," but research in emotional neuroscience shows that emotions involve more than just the amygdala. In particular, developing the prefrontal cortex during adolescence can make emotion regulation more difficult, leaving young people especially vulnerable to emotional manipulation. Although studies [2, 6] summarized research on the impact of perception and presence on emotional reactions in VR, they focused on single users, which is far from the emotional behaviors in the multifaceted social interactions in the Metaverse. The following two examples depict emotionally hijacked scenarios through social interactions in the Metaverse.

- Impulsive-Compulsive Buying Disorder (ICBD): Emotion-based impulsivity can cause individuals to make rash decisions due to outside stressors. Those with Impulsive-Compulsive Buying Disorder (ICBD) [1] may be particularly vulnerable to developing the disorder through Metaverse's euphoria.
- State-dependent Memory: Perpetrators can exploit the anonymity of avatars in the Metaverse and use targeted profiling to manipulate users. Perpetrators can pretend to be someone the victim knows, triggering state-dependent memory [8] and causing victims to remember more information.

As the Metaverse becomes increasingly popular, people become more susceptible to emotional manipulation in such a virtual environment. In this work, we will conduct social interaction experiments in VR to quantify the strength of emotions from the EEG signals measured from the brain to understand better human emotions evoked by realistic situations in the Metaverse. The proposed neurophysiological approach offers a new way to assess emotional reactions in the Metaverse, as it provides objective human response data, captures responses synchronized with specific virtual cues in real-time, and evaluates implicit affective processes that were previously inaccessible. In addition, this approach creates a comprehensive emotional perception assessment framework, reveals complex relationships between environmental cues and emotional reactions, and identifies the influence of emotional cues and their representations to inform advanced research in the Metaverse.